\documentclass[11pt,twoside]{article}

\usepackage{asp2004, epsf, psfig, graphicx, lscape}

\begin{document}
\title{Evidences on Secular Dynamical Evolution of Detached Active Binary 
Orbits and Contact Binary Formation}
\author{Z. Eker}  
\affil{\c{C}anakkale University Observatory, 17100 \c{C}anakkale, Turkey}    
\author{O. Demircan}
\affil{\c{C}anakkale University Observatory, 17100 \c{C}anakkale, Turkey} 
\author{S. Bilir}
\affil{Istanbul University, Faculty of Science, Department of Astronomy and Space
      Sciences, 34119, Istanbul, Turkey}
\author{Y. Karata\c{s}}
\affil{Istanbul University, Faculty of Science, Department of Astronomy and Space
      Sciences, 34119, Istanbul, Turkey}

\begin{abstract}
Evidence of secular dynamical evolution for detached active binary orbits are 
presented. First order decreasing rates of orbital angular momentum (OAM), systemic 
mass ($M=M_{1}+M_{2}$) and orbital period of detached active binaries have been 
determined as $\dot J/J = -3.48 \times 10^{-10}$yr$^{-1}$, $\dot M/M = -1.30 \times 
10^{-10}$yr$^{-1}$ and $\dot P/P = -3.96\times 10^{-10}$yr$^{-1}$ from the kinematical 
ages of 62 field detached systems. The ratio of $d \log J/ d \log M = 2.68$ implies 
that either there are mechanisms which amplify AM loss $\delta=2.68$ times with 
respect to isotropic AM loss of hypothetical isotropic winds or there exist external 
causes contributing AM loss in order to produce this mean rate of decrease for orbital 
periods. Various decreasing rates of OAM ($d \log J / dt$) and systemic mass 
($d \log M/ dt$) determine various speeds of dynamical evolutions towards a contact 
configuration. According to average dynamical evolution with $\delta = 2.68$, the 
fraction of 11, 23 and 39 per cent of current detached sample is expected to be 
contact system within 2, 4 and 6 Gyr respectively.
\end{abstract}

\section{Introduction}
Observational data and theory of contact binaries revised extensively by \cite{Mohnachi81}, 
\cite{Vilhu81} and \cite{Rucinski82}. According to \cite{Rucinski86} the most promising 
mechanism of formation for contact binaries involves the orbital angular momentum loss 
(AML) and the resulting orbital decay of detached but close synchronized binaries. AML by 
magnetic breaking (\citeauthor{Schatzman59} \citeyear{Schatzman59}, \citeauthor{Kraft67} 
\citeyear{Kraft67}, \citeauthor{Mestel68} \citeyear{Mestel68}) became popular especially 
after \cite{Skumanich72} who presented observational evidence of decaying rotation rates 
for single stars. Magnetic breaking and tidal locking have been considered as a main route to form 
W~UMa-type contact binaries from the systems initially detached (cf. \citeauthor{Huang66} 
\citeyear{Huang66}, \citeauthor{Okamoto70} \citeyear{Okamoto70}, \citeauthor{van'tVeer79} 
\citeyear{van'tVeer79}, \citeauthor{Vilhu80} \citeyear{Vilhu80}, \citeauthor{Mestel84} 
\citeyear{Mestel84}, \citeauthor{Guinan88} \citeyear{Guinan88}, \citeauthor{Maceroni91} 
\citeyear{Maceroni91}, \citeauthor{Stepien95} \citeyear{Stepien95}, \citeauthor{Demircan99} 
\citeyear{Demircan99}).

Orbital period evolutions and time scale for forming contact systems from detached 
progenitors are predicted differently among the authors above. Since tidal locking is 
more effective at short periods, secular orbital period decreases were estimated slow at 
the beginning. Only the binaries of few days orbital periods were predicted to become 
contact systems within the order of $\sim10^9$ years. Shrinking orbits and related orbital 
period decreases, unfortunately, are not detectable on commonly used O-C diagrams formed 
by eclipse times due to: (1) short time-span covered by existing O-C data (at most 100 
years), (2) large scattering of unevenly distributed O-C data, (3) existence of complicated 
larger amplitude shorter time-scale fluctuations by many different effects such as mass 
transfer, third companion and/or magnetic cycles etc. (cf. \citeauthor{Demircan00} 
\citeyear{Demircan00},\citeyear{Demircan02}; \citeauthor{Kreiner01} \citeyear{Kreiner01}). 
Observed period decreases on O-C diagrams, thus, cannot be counted as observational evidence 
for the secular orbital period decreases. 

There exists an opposing theory which predicts different scenario of contact binary formation 
by a fission process \citep{Roxburgh66} enabling contact binaries to form at the end of the 
pre-main sequence contraction. Considering the brief lifetime of contact stages 
($0.1 < t_{contact} < 1$ Gyr) estimated by \cite{Guinan88}, this theory would fail to 
produce older population of W UMa binaries unless AM is conserved. However, conservation 
of AM would refute formation of contact systems from the detached progenitors.

Debate on the formation mechanism continues. Low space density of contact binaries (0.2\% 
in the solar neighborhood, \citeauthor{Rucinski02} \citeyear{Rucinski02}, \citeyear{Rucinski06})
implies limited lifetimes for contact stages and favors the formation mechanism from the 
detached progenitors. However, if tidal locking is efficient only  at orbital periods 
comparable ($\sim$1 days) to contact systems, then the 0.2\% space density becomes too much 
that ``contact binaries appear out of nowhere" \citep{Paczynski06}. Other mechanisms of OAM 
loss besides tidal locking must occur to account this observed space density in the solar 
neighborhood. 

This presentation aims to summarize the evidences of secular dynamical evolution for 
detached binary orbits recently obtained by us from the kinematics of chromospherically 
active binaries \citep{Karatas04} and from the kinematics of W UMa systems \citep{Bilir05}. 
After a brief summary of orbital dynamics, mean dynamical evolution according to 
\cite{Demircan06} will be described on  $\log J-\log P$, $\log M-\log P$ and $\log J-\log M$ 
diagrams.

\section{Evidence of secular decrease of orbital periods and masses} First, \cite{Guinan88} 
estimated a kinematical age of 8-10 Gyr for W UMa systems from their galactic space velocity 
dispersions. Theory of formation from detached progenitors appears consistent even if this 
age is compared to an earlier estimate of $\sim5$ Gyr kinematical age \citep{Eker92} of 
possible progenitors; the chromospherically active binaries (CAB).
Being comparable to nuclear time-scale, the 8-10 Gyr age of W UMa systems 
implies that contact systems either have been retaining their original AM according to 
\cite{Roxburgh66} or have been formed from detached binaries if orbits are unstable 
against AM loss. However, according to the space velocities and the dispersions 
by \cite{Aslan99}, W UMa binaries are not older than RS CVn systems. 

Increased size of the samples (CAB \& W UMa) together with greatly improved astrometric data 
(parallaxes \& proper motions) by Hipparcos \citep{Perryman97} motivated us to reanalyze the 
problem once more. First, identifying possible members of young moving groups (MG) from 237 
CAB binaries, we had spitted them into two sub samples of kinematically young ($<0.95$ Gyr, 
N=95) and older field (3.86 Gyr, N=142) systems.

MGs are kinematically coherent groups of stars that share a common origin. \cite{Eggen94} 
defined a super cluster of stars gravitationally unbound in the solar neighborhood, but 
having a same kinematics while occupying the extended regions in the Galaxy. Therefore a 
MG, unlike well known open clusters covering only a limited sky, can be observed at all 
directions. Kinematical criteria originally defined by \cite{Eggen58a, Eggen58b, Eggen89, 
Eggen95} for determining possible members of the best-documented MGs are summarized by 
\cite{Montes01a, Montes01b}. The basic idea is that a test star's space velocity vector 
must be equal and parallel, or at least with deviations smaller than the pre-determined 
limits, to the space velocity vector of a MG. The ages of MGs are known as open cluster 
ages from the main-sequence turn-off point. Ages of MG groups considered in this study 
are given in Table 1 together with kinematical parameters to identify them.

\begin{table}
{\tiny
\caption{Parameters of best documented moving groups.} 
\center
\begin{tabular}{lcccccc}
\hline
Name &        Age &  (U, V, W) &    $V_{T}$ &  C.P. \\
     &      (Myr) &     (km/s) &     (km/s) &   ($\alpha^{h}$, $\delta^{o}$)     \\
\hline
Local Association &   20 -- 150 & (-11.6,-21.0,-11.4) &     26.5 & (5.98,-35.15) \\
(Pleiades, a Per, M34, &            &            &            &                  \\
$\delta$ Lyr, NGC 2516, IC2602) &            &            &            &         \\
IC 2391 Supercluster &    35 -- 55 & (-20.6,-15.7,-9.1) & 27.4 & (5.82,-12.44) & \\
 (IC 2391) &            &            &            &                              \\
 Castor MG &        200 & (-10.7,-8.0,-9.7) &       16.5 & (4.75,-18.44)         \\
Ursa Major Group &        300 & (14.9,1.0,-10.7) &       18.4 & (20.55,-38.10)   \\
(Sirius Supercluster) &            &            &            &                   \\
Hyades Supercluster &        600 & (-39.7,-17.7,-2.4) &       43.5 & (6.40,6.50) \\
(Hyades, Praesepe) &            &            &            &                      \\
\hline
\end{tabular} 
} 
\end{table} 

Dispersions of CAB and W UMa stars on the $U-V$ diagrams are compared 
in Fig. 1, where possible MG members are also shown. $U$ \& $V$ are space velocity  
components towards the Galactic center and in the direction of Galactic rotation. 
>From the galactic space velocity dispersions, 3.86 and 5.47 Gyr of kinematical ages 
were assigned to field CABs and W UMa's (\citeauthor{Karatas04} \citeyear{Karatas04}, 
\citeauthor{Bilir05} \citeyear{Bilir05}). Although being smaller than earlier 
determinations, these ages too do not show a conflict with the theory of formation 
from detached progenitors. Moreover, the difference (1.61 Gyr) between these ages 
was interpreted as a mean lifetime of contact stages by \cite{Bilir05}. 

\begin{figure}
\begin{center}
\includegraphics[width=130mm, height=124.6mm]{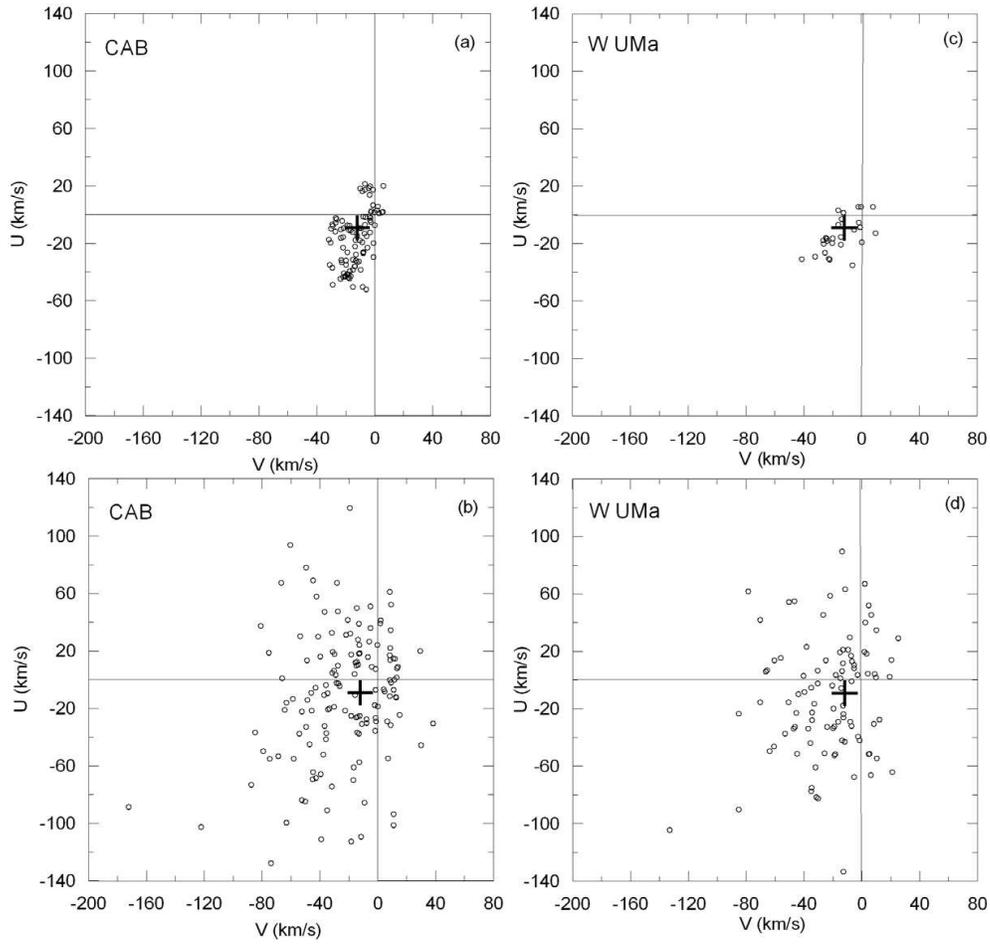}
\caption[] {CAB (left), W UMa (right), Possible MG members (upper) are removed from 
the samples, then what left are called field systems (below).} 
\end{center}
\end {figure}

\begin{figure}
\begin{center}
\includegraphics[angle=0, width=130mm, height=41mm]{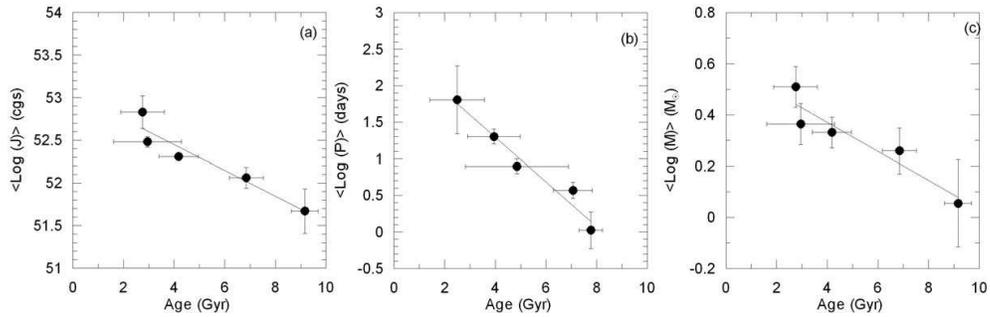}
\caption[] {Age dependent variations of OAM ($J$), period ($P$) and mass ($M$).} 
\end{center}
\end {figure}

Formed by kinematical criteria, young groups allowed us to compare period and 
total mass histograms as well as other physical parameters between a young group 
and corresponding field systems in order to search further evidences of OAM loss. 
Since kinematical criteria determine only possible members for a MG, wrong 
identifications are always possible since the space velocity of a star coincidently 
may imply a membership. We believe number of wrong identifications are small and 
thus unable to spoil statistics that evidences of dynamical evolution from 
detached to contact stages could still be found. 

Being much younger than field systems, histograms of MG group CAB could be taken as 
initial distributions of field CAB. Consequently, it becomes clear that smooth shaped 
initial distribution (see Fig. 6 of \citeauthor{Karatas04} \citeyear{Karatas04}) is 
changed to a distribution with a peak at 10 days indicating the relative number of 
longer and shorter period systems were decreased. Secular OAM loss forcing orbital 
periods to decrease could explain the reduction of the relative number of systems 
with periods longer than 10 days. On the other hand, period decrease and radius 
increase as a result of dynamical and nuclear evolutions must have changed the 
missing short period systems into contact or semi-contact form thus they are no 
longer in the list of current CABs. We think the difference of period histograms 
between the MG group and the field CABs presents clear hints of OAM loss and 
secular decrease of orbital periods. 

It is not as clear, but similar kind of trends are also 
noticeable in the total mass ($M = M_{1} + M_{2}$) histograms 
(see Fig. 8 of \citeauthor{Karatas04} \citeyear{Karatas04}). 
The tails rather than peaks support the prediction of mass decrease as a consequence 
of mass loss which carries away OAM of active binaries. The gradual decrease of the 
high mass tail of young group is changed to a sharper decrease in the older group. 
Similarly sharp decrease towards the less massive systems changed to rather a gradual 
decrease in the older population. Heterogeneity of samples containing giants (G), 
sub-giants (SG) and main-sequence (MS) systems and the evolution into contact or 
semi-contact configurations complicates the histograms and makes the interpretation 
of the peaks more difficult. See \cite{Karatas04} for the details and  further 
interpretations about sub-groups discriminating G, SG and MS systems. 

Increase of kinematical ages towards the short period systems (see Table 5 of 
\citeauthor{Karatas04} \citeyear{Karatas04}) could be taken as the clearest evidence 
of orbital period decrease as a consequence of OAM loss. Dynamical evolution has 
been further quantized with mean decreasing rates of OAM, total masses and periods 
from the kinematical ages of carefully selected detached CABs with most reliable 
physical parameters by \citeauthor{Demircan06} (\citeyear{Demircan06}), who preferred 
to form sub-samples according to OAM ranges. Mean OAM ($J$), period ($P$) and mass 
($M$) of these sub groups versus the kinematical ages are displayed in Fig. 2. The 
decreases of $J$, $P$ and $M$ by age are obvious. As a first approximation linear 
lines were fitted by the least squares method and the inclinations were found to be 
$-1.51\times  10^{-10}$ yr$^{-1}$, $-1.72\times 10^{-10}$ yr$^{-1}$, and 
$-5.65\times 10^{-11}$ yr$^{-1}$.
  
A linear change in the logarithmic scale implies a constant rate of change. 
Consequently, the mean relative decreasing rates $\dot J/J=-3.48\times10^{-10}$yr$^{-1}$, 
$\dot P/P=-3.96\times 10^{-10}$yr$^{-1}$and $\dot M/M=-1.30 \times 10^{-10}$yr$^{-1}$ were 
determined from kinematics to represent a mean dynamical evolution have been occurring 
among the detached CAB stars in the solar neighborhood.

Similar histograms and tables were also produced for W UMa stars. However, 
similar kind of interpretations are not possible for them because lifetime 
in the contact stage is too short, e.g. $<1$ Gyr \citep{Guinan88} or $\sim 1.6$ 
Gyr \citep{Bilir05}. The MG group systems would disappear before reaching to an 
age of even the youngest group of field contacts with a mean kinematical age 
3.21 Gyrs (See Table 6 of \citeauthor{Bilir05} \citeyear{Bilir05}). The four 
sub-groups formed from 97 field contact binaries according to orbital period 
ranges (ages 3.21, 3.51, 7.14 and 8.89 Gyrs) dominate over 27 systems which are 
possible MG members with ages less than 0.6 Gyrs. This fact implies that 
pre-contact dynamical evolution must exist dominantly among W UMa stars. 
Nevertheless, being out of the ordinary, direct formation at the end of pre-main 
sequence contraction by fission 
process \citep{Roxburgh66} must also be occurring because ages of MG groups 
($< 0.6$ Gyrs) give no space for pre-contact detached phases. Due to complexities 
of various lengths of pre-contact stages, it is not possible to evaluate the total 
mass and the period histograms for W UMa systems in a similar manner as CAB systems.

\section{Orbital dynamics}
The most basic definition of OAM ($J$) can be given as

\begin{eqnarray}
J={M_{1}M_{2}\overwithdelims()M_{1} + M_{2}}{a^2}\Omega=
{q \overwithdelims()(1 + q)^{2}}{M a^2}\Omega.
\end{eqnarray}
where $I={M_{1}M_{2}\overwithdelims()M_{1} + M_{2}}{a^2}={q \overwithdelims()(1 + q)^{2}}{M a^2}$
is moment of inertia and $\Omega = 2\pi/P$ is angular speed for an orbital motion, thus $J=I\Omega$.

OAM ($J$) is needed dynamically to keep the orbital motion. Therefore, $J$  and $M$ are physical quantities
which determine a unique period ($P$) and a unique size for the orbit as

\begin{eqnarray}
P={(1+q)^{6} \over q^3} {2\pi \over G^2}{ J^3\over M^5},\qquad\qquad a={(1+q)^{4} \over Gq^2} { J^2\over M^3},
\end{eqnarray} 
where the mass ratio ($q = M_{2}/M_{1} < 1$) can be considered as an auxiliary 
parameter used in the definition of $J$. The size $a=a_1+a_2$ represents the 
semi--major axis of a  relative orbit of one star around the other. Stability 
of an orbit ($dP=0$, $da=0$) requires OAM and mass to be constant ($dJ=0$, 
$dM=0$) provided with no mass transfer ($dq=0$). If there is no mass transfer, 
which must be true for detached binaries, it is obvious that OAM loss will 
cause an orbit to reduce its period and size. On the contrary, mass loss has 
an affect of increasing the period and size. Logarithmic derivatives of (2) 
give possible relative changes as

\begin{eqnarray}
{dP \over P} = -3 {{1-q}\over{1+q}} {dq\over q} +3 {dJ \over J}-5{dM\over M},\qquad\qquad {da\over a} = -2 {{1-q}\over{1+q}} {dq\over q} +2 {dJ \over J}-3{dM\over M}.
\end{eqnarray} 
Because $M$ has higher power than $J$, the affect of mass loss would dominate. 
For example, in the case of same relative changes of OAM and mass (isotropic 
stellar winds, if $dq=0$), $dJ/J=dM/M$ according to (1), then

\begin{eqnarray}
{dP \over P} = -2{dJ \over J}=-2{dM\over M},\qquad\qquad {da\over a}=-{dJ \over J}=-{dM\over M}.
\end{eqnarray} 
which means mass loss and corresponding OAM loss will have a net effect on the 
orbit to increase both the period and the size. However, there could be additional 
causes to increase relative OAM loss, e.g. OAM loss of  gravity waves, or stellar 
encounters in the galactic space, or a third body in an eccentric orbit around 
the binary system, or existence of an amplification mechanism as in some tidally 
locked binaries, in which tidal interactions transfer OAM to spinning components and  
AM is lost at the Alfven radius. After considering all 
possibilities, one has to compare the grant total relative OAM loss to the 
relative mass loss which could be expressed by a parameter $\delta$ defined as 

\begin{eqnarray}
{\delta} = {({dJ \over J}) / ({dM \over M})},
\end{eqnarray}
which can be called dynamical parameter because dynamical respond of the orbit 
depends on the value of $\delta$. Using this definition of $\delta$, equation (3) becomes

\begin{eqnarray}
{dP \over P} = (3-{5\over \delta}){dJ \over J}=(3\delta -5){dM\over M},\qquad\qquad {da\over a} = (2-{3\over \delta}){dJ \over J}=(2\delta -3){dM\over M},
\end{eqnarray}
if $dq/q$ is neglected. Mass ratio 
change could be zero just because relative mass loss of components would be 
equal. Even if $dq/q \neq 0$, the term $(1-q)/(1+q)$ could be very small 
especially for high mass ratio ($q\sim1$) systems. Thus, ignoring it in the 
first approximation is acceptable. 
 
For decreasing an orbital period, $\delta>5/3$ is required. But, 
$\delta > 3/2$ is sufficient to shrink an orbit. If $3/2<\delta<5/3$, orbital size  
decreases despite period is increasing. The size and the period of an orbit both 
increase if $\delta < 3/2$.
  
Using the mean decreasing rates of OAM and mass from the kinematical ages of 
detached CAB stars, a mean value for the dynamical parameter ($\bar{\delta}$) 
can be estimated for them as
\begin{eqnarray}
{\bar {\delta}} = {{dJ \over J} \over {dM \over M}} = {{dJ \over Jdt} \over {dM \over Mdt}} = {-3.48 \times 10^{-10} \over -1.30 \times 10^{-10}}=2.68 
\end{eqnarray}
in the solar neighborhood.

\section{Mean dynamical evolution on diagrams}
\subsection{The $\log J-–\log P$ diagram}
CAB and W UMa stars with available OAM are plotted on a $\log J–-\log P$ 
diagram (Fig. 3). The CAB stars containing giants, sub-giants, and main sequence 
and A \& W type W UMa are indicated. W UMa stars are located at the lower left. 
Having larger masses and orbital periods, the CAB systems display a band like 
distribution on the right of W UMa stars as elongated from lower left to upper 
right. The constant total mass lines are computed with $q=0.88$, which is the median 
value of mass ratios of CAB stars, using      
\begin{eqnarray}
J={q\over (1+q)^2}\sqrt[3]{{G^2 \over 2\pi}M^5P}
\end{eqnarray}
at which $P$ is varied while chosen $M$ values are fixed. 

\begin{figure}
\begin{center}
\includegraphics[angle=0, width=100mm, height=56.5mm]{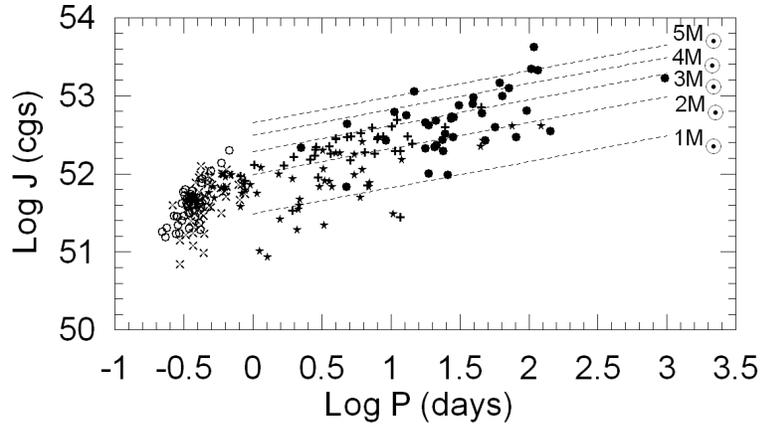}
\caption[] {CAB stars containing giants ($\bullet$), subgiants ($+$), main-sequence 
($\star$); W UMa stars A type ($\times$), W type ($\circ$). Constant mass 
($M=M_1+M_2$) lines ($--$) are computed with $q=0.88$ for CAB systems.}
\end{center}
\end {figure}

\begin{figure}
\center
\includegraphics[angle=0, width=80mm, height=77.44mm]{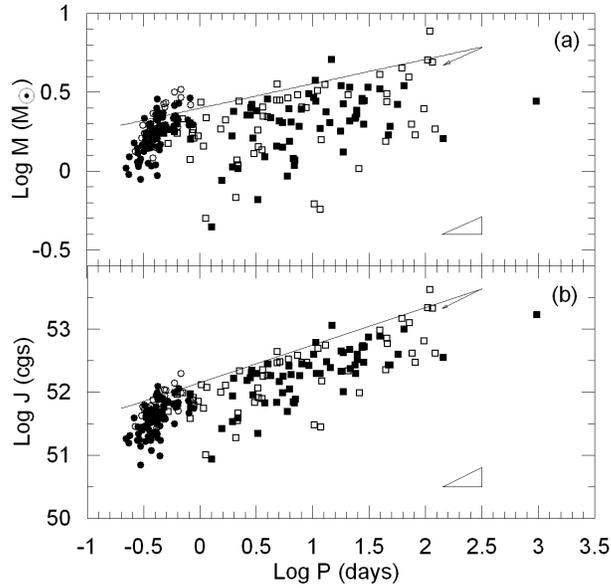}
\caption{Upper boundary in (a) could be translated to (b) analytically. 
CAB (\sq), W UMa ($\circ$); MG (empty), field (filled). Young and old CABs are randomly mixed. 
Mean dynamical evolution (arrow), mean decreases (right sides of triangles) for 2 Gyrs.}
\end{figure}

Well defined smooth upper boundary of CAB systems appears as if tracing the path 
of dynamical evolution for them. OAM loss, mass loss and associated orbital period 
decrease would  move a system from upper right to lower left as if parallel to the 
upper boundary. Finally, some systems would enter in the region of W UMa stars. 
However, this description is not quite correct according to the mean dynamical 
evolution with $\bar \delta = 2.68$, which is marked by an arrow at the upper end 
of the upper boundary in Fig. 4b. The magnitude of the arrow symbolizes an interval 
of 2 Gyrs. The right sides of the triangle in the lower right show the amount of 
OAM loss and corresponding orbital period decrease.

Fig. 4b plotted with symbols to indicate kinematically young (MG) and 
old (field) stars. The random distribution of the young and old CAB systems in 
implies that a detached system may start its dynamical evolution anywhere 
on the diagram in the region of CAB stars. So, it is not possible to distinguish 
younger and older systems according to their location on these diagrams unlike 
the nuclear evolution on H-R diagram which has a well defined starting point on 
the Zero Age Main Sequence (ZAMS). It is well known that the location of a star 
on H-R diagram indicates its stage of evolution as well as its age. 

\subsection{The $\log M-– \log P$ diagram}
The $\log M-–\log P$ diagram resembles a similar type distribution as the 
$\log J-–\log P$ diagram. First, the upper boundary of CAB stars on the $
\log M-–\log P$ diagram (Fig. 4a) was eye estimated and digitized by computer. 
Then using its numerical $M$, $P$ points together with $q=0.88$, the solid line 
on the $\log J-–\log P$ diagram was computed by (8). Since the computed line 
fits even better to the CAB upper boundary of $\log J-–\log P$ distribution, 
the upper boundaries of both diagrams in Fig. 4 are not independent and appears 
to be determined by the upper mass limits of orbital periods. Non-existence of 
CAB stars above those upper boundaries could be related to binary formation 
mechanism since initial Roche lobes may put limits on the masses of forming 
binaries. The binaries above, if they exist, they are not chromospherically 
active because being brighter they would have been easier to be noticed as 
CAB stars, then upper boundaries on those diagrams mark only the upper mass 
limit for a binary of a given period to have choromosphric activity.

Moreover, If period decrease occurs because of OAM loss but without mass loss 
and mass transfer, a dynamical evolution would follow a path parallel to the 
constant mass lines in Fig. 3, which would carry binaries into the empty region. 
This requires for a CAB system to seize chromospheric activity if it moves into 
the empty region by such a process. Therefore, it is most likely the CAB upper 
boundary on $\log J-–\log P$ diagram represent a dynamical evolution with a 
minimum mass loss. Nevertheless, mean decreases determined from the kinematical 
ages of CAB stars shown by the right sides of the triangles in Fig. 4 indicates 
that the direction of mean dynamical evolution neither is parallel nor towards 
the empty region as it is shown by the arrows at the upper ends of the upper 
boundaries in Fig. 4. 

\subsection{The $\log J–-\log M$ diagram}
Because $J$ and $M$ are basic quantities to determine $P$ \& $a$ and because 
OAM loss \& mass loss are parameters controlling the magnitude and direction 
of dynamical evolution, the $\log J-–\log M$ diagram is a natural choice 
to study dynamical evolution of orbits. Once, Fig. 5 is 
produced, a sharp separation between the detached and contact systems stroke 
to our attention. Goodness of separation is out striking that despite crowding 
along the border, there are only two systems (OO Aql, $\delta$ Cap) on the 
wrong side, which could be due to a wrong identification of the state of being 
contact or just because of observational errors. Separation of such a degree 
does not occur on the diagrams discussed before.

Marking several positions on the borderline between CABs and W UMas following 
quadratic equation was produced.
\begin{eqnarray}
\log J=0.522(\log M)^{2}+1.664(\log M)+51.315,
\end{eqnarray}  
where $M$ is in solar units and $J$ is in cgs. Physical significance of this 
line is that it marks the maximum OAM for a contact system to survive. 
It is like in single stars, spin AM has to be less than a certain value 
otherwise gravity cannot hold stellar mass together. It is same for contact 
binaries, if OAM is more than the value computed above, 
the contact configuration brakes. CAB systems (RT Lac, AR Mon, $\epsilon$ UMi, 
RV Lib, BH CVn), which were eliminated from the list of \cite{Demircan06} when 
the mean decreasing rates of $J$, $M$ and $P$ were 
determined (since they are filling or about to fill one of the Roche lobes 
that mass transfer possibly occurring in them), are marked on Fig. 6. Since 
those systems are not close to the contact border and scattered randomly all over in 
the detached region, the state of being detached and 
semi-detached must be same physically. 

\begin{figure}
\center
\resizebox{130mm}{62mm}{\includegraphics*{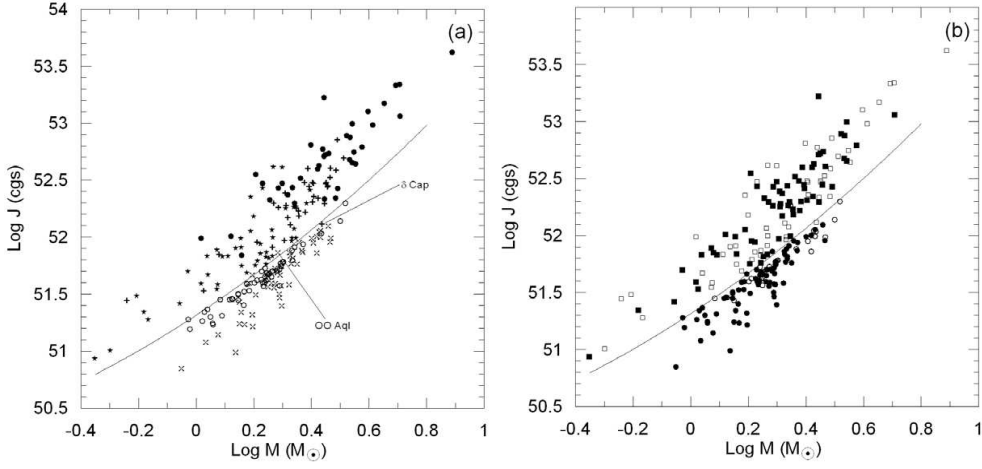}} 
\caption{Well defined borderline sharply separates detached and contact systems. 
(a) Symbols are like Fig. 3, (b) Symbols are like Fig. 4. Young and old CABs are 
randomly mixed.}
\end{figure}

A detached system must lose OAM to go into the 
region of contacts in order to be a contact binary. This possibility, however, 
very much depends on the position on the diagram together with the direction 
and the speed of the dynamical evolution, which could be very 
different from one system to another. The $\log J-–\log M$ diagram too gives no 
clue on individual dynamical evolutions since there is no information about  
initial positions. Random mix of young (MG) and old (field) systems (Fig. 5b) 
indicates that a same position could be belonging to both a young and an old 
system.

\begin{figure}
\center
\resizebox{60mm}{57.5mm}{\includegraphics*{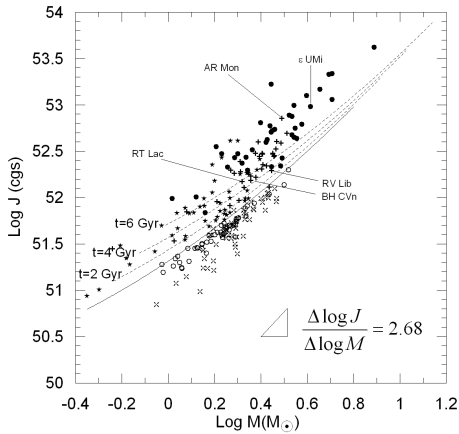}} 
\caption{Loci of equal times to reach at the contact border. Mean dynamical 
evolution (hypothenuse) and mean decreases (right sides) for 2 Gyr.}
\end{figure} 

Nevertheless, the contact border (eq. 9) could be helpful to determine contact 
binary candidates. The amount of mean losses ($\Delta J$ and $\Delta M$) corresponding 
to 2, 4 and 6 Gyr are subtracted from the $J$ and $M$ values of the border. That is, 
the contact border is shifted accordingly. Dotted lines in Fig. 6 represent 
shifted borders that the systems between a dotted line and the contact border are 
the ones which are predicted to be contact systems within the time intervals  
indicated if their nuclear evolutions permit them to live untill the predicted times. 
After counting, it becomes clear that 11, 23 and 39 per cent of the 
current sample of CABs could pass over the contact border within the 
next 2, 4 and 6 Gyr according to the mean dynamical evolution with 
$\bar \delta=2.68$.    

\subsection{Period and size evolution of orbits}

It is possible to draw constant period lines using (8). On the other hand, 

\begin{eqnarray}
J={q\over (1+q)^2}\sqrt{GM^3a}
\end{eqnarray} 
can be used to compute constant orbital size lines similarly. From the statistics 
of present samples, median values $q=0.88$ and $q=0.39$ are found to represent  
CAB and W UMa stars. The constant orbital period and orbital size lines run 
almost parallel to the contact border and both $P \& a$ values decrease towards 
it (Fig. 7). Further decrease into the region of contacts is also clear. 

\begin{figure}
\center
\resizebox{130mm}{62.10mm}{\includegraphics*{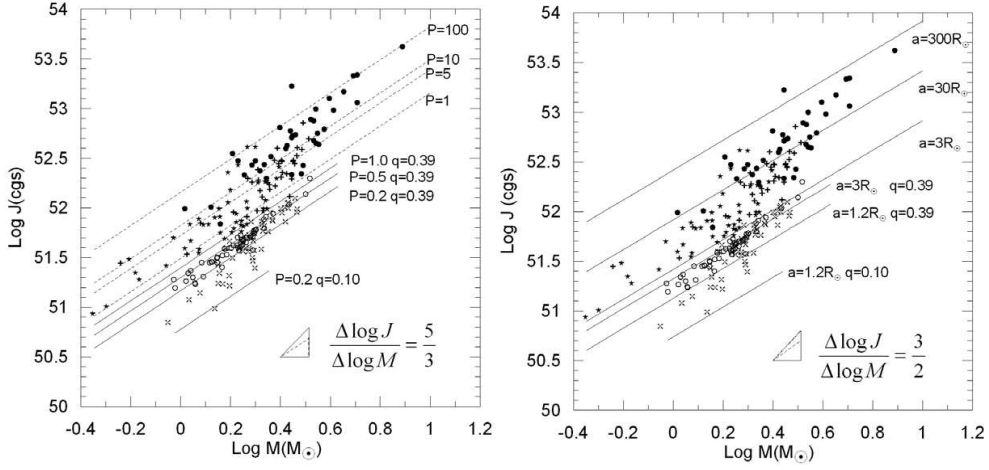}}
\caption{Period and size evolutions of orbits. Mean (hypotenuse) dynamical 
evolution and corresponding decreases (right sides of triangles). Evolution with 
$\delta=5/3$ keeps $P$ constant but $\delta=3/2$ keeps a constant.}
\end{figure}

Constant period and constant size lines are sensitive to small $q$ values. One 
can feel the sensitivity by comparing $P=1$ day (or $a=3R_{\odot}$) lines of 
$q=0.88$ and $q=0.39$, also $P=0.2$ days (or $a= 1.2R_{\odot}$) computed with 
$q=0.39$ and $q=0.1$. Decreasing periods (or sizes) towards the lower right is 
deceptive as if evolution to contact stage is occurring from upper left to lower 
right which is impossible since such an evolution requires a mass gain. Mass 
loss, however, changes the direction from vertically down (OAM loss only) towards 
to the lower left (if OAM and mass both are lost). Because the affect of mass 
loss dominates over OAM loss (eq. 3), there are lower limits; one for orbital 
periods and one for the orbital sizes. Both limits are indicated by the dotted 
lines in the triangles and the corresponding numerical values ($d \log J/ d \log M$) in 
Fig. 7. Any dynamical evolution with a $\delta$ smaller than those limits 
indicates an increase rather than a decrease on both $P \& a$. 

\section{Conclusions}

Since $\bar \delta = 2.68$ for detached CAB in the solar neighbourhood is greater 
than both limits, (5/3 for $P$, 3/2 for $a$) orbits are shrinking and periods are 
decreasing. A well defined borderline sharply separating detached and contact systems 
are discovered empirically on $\log J-–\log M$ diagram. It is possible for a detached 
system to pass over the contact border and become a contact system by OAM loss. Not 
only $P\sim1$ day period detached systems, but also some $P\sim10$ days period 
systems can pass over this border according to the constant period lines in Fig. 7 
and loci of equal times to reach at the contact border in Fig. 6. Significant number 
of current CABs (39 per cent) were found as canditates to pass over the contact border 
within the next 6 Gyrs according to their positions on the $\log J-\log M$ diagram.
Nevertheless, contact binary 
formation from detached progenitors is not the only mechanism to form contact systems. 
Although, it is rare, direct formation at the end of the pre-main sequence evolution 
must also be occurring. Statistical studies are encouraged to include these new findings if one can still 
say ``contact binaries appear out of nowhere" 
while they are already known to be rare systems \citep{Rucinski02, Rucinski06}.

Acknowledgements: This work has been partially supported by COMU BAP 2005/108, Research 
Fund of the University of Istanbul (244/23082004) and TUBITAK 104T508.


\begin{thebibliography}{}
{\small
\bibitem[Aslan et al.(1999)]{Aslan99}
Aslan, Z., \H{O}zdemir, T., Ye\c{s}ilyaprak, C., \.Iskender, A., 1999, 
Tr. J. of Physics, 23, 45

\bibitem[Bilir et al.(2005)]{Bilir05}
Bilir, S., Karata\c{s}, Y., Demircan, O., Eker, Z., 2005, MNRAS, 357, 497

\bibitem[Demircan(1999)]{Demircan99}
Demircan, O., 1999, Tr. J. Phys., 23, 425

\bibitem[Demircan(2000)]{Demircan00}
Demircan, O., 2000, in Variable Stars as Essential Astrophysical Tools, ed. 
C \.{I}bano\u{g}lu. Dordrecht; Boston (NATO Science Series. Series C.
Mathematical and Physical Sciences; Vol 544), p.615

\bibitem[Demircan(2002)]{Demircan02}
Demircan, O., 2002, in The Royal Road to Stars, ed. O Demircan, E Budding. 
Publications of COMU, \c{C}anakkale, Turkey, p. 130

\bibitem[Demircan et al.(2006)]{Demircan06}
Demircan, O., Eker, Z., Karata\c{s}, Y., Bilir, S., 2006, MNRAS, 366, 1511

\bibitem[Eggen(1958a)]{Eggen58a}
Eggen, O.J., 1958a, MNRAS, 118, 65

\bibitem[Eggen(1958b)]{Eggen58b}
Eggen, O.J., 1958b, MNRAS, 118, 154

\bibitem[Eggen(1989)]{Eggen89}
Eggen, O.J., 1989, PASP, 101, 366

\bibitem[Eggen(1994)]{Eggen94}
Eggen, O.J., 1994, in Morrison L. V., Gilmore G., eds, Galactic and Solar 
System Optical Astrometry. Cambridge Univ. Press, Cambridge, p.191

\bibitem[Eggen(1995)]{Eggen95}
Eggen, O. J., 1995, AJ, 110, 2862

\bibitem[Eker(1992)]{Eker92}
Eker, Z., 1992, ApJS, 79, 481

\bibitem[Huang(1966)]{Huang66}
Huang, S.S., 1966, ARA\&A, 4, 35

\bibitem[Guinan \& Bradstreet(1988)]{Guinan88}
Guinan, E.F., Bradstreet, D.H., 1988, in Dupree A.K., Lago M.T., eds, 
Formation and Evolution of Low Mass Stars, Kluwer, Dordrecht, p. 345 

\bibitem[Karata\c{s} et al.(2004)]{Karatas04}
Karata\c{s}, Y., Bilir, S., Eker, Z., Demircan, O., 2004, MNRAS, 349, 1069

\bibitem[Kreiner, Kim \& Nha(2001)]{Kreiner01}
Kreiner, J.M., Kim, C., Nha, II-Seung, 2001, on Atlas of O–C Diagrams of 
Eclipsing binary Stars, Poland: Wydawnictwo Noukowe Akademii

\bibitem[Kraft(1967)]{Kraft67}
Kraft, R.P., 1967, ApJ, 150, 551

\bibitem[Maceroni \& van't Veer(1991)]{Maceroni91}
Maceroni, C., van't Veer, F., 1991, A\&A, 246, 91

\bibitem[Mestel(1968)]{Mestel68}
Mestel, L., 1968, MNRAS, 138, 359 

\bibitem[Mestel(1984)]{Mestel84}
Mestel, L., 1984, in  S.L. Baliunas and L. Hartmann, eds, the Third Cambridge 
Workshop Cool Stars, Stellar Systems, and the Sun, Lecture Notes in Physics, 
Vol. 193, Springer-Verlag, Berlin, Heidelberg, New York, p. 49 

\bibitem[Mohnachi(1981)]{Mohnachi81}
Mohnachi, S.W., 1981, ApJ, 245, 650

\bibitem[Montes et al.(2001a)]{Montes01a}
Montes, D., Lopez-Santiago, J., Galvez, M.C., Fernandez-Figueroa, M.J., 
De Castro, E., Cornide, M., 2001a, MNRAS, 328, 45

\bibitem[Montes et al.(2001b)]{Montes01b}
Montes, D., Fernandez-Figueroa, M. J., De Castro, E., Cornide, M., Latorre, 
A., 2001b, 11th Cambridge Workshop on Cool Stars, Stellar Systems and the 
Sun, ASP Conference Proceedings, Vol. 223. Edited by Ramon J. Garcia Lopez, 
Rafael Rebolo, \& Maria Rosa Zapaterio Osorio. San Francisco: Astronomical 
Society of the Pacific, p.1477

\bibitem[Okamoto \& Sato(1970)]{Okamoto70}
Okamoto, I., Sato, K., 1970, PASJ, 22, 317 

\bibitem[Paczynski et al.(2006)]{Paczynski06}
Paczynski, B., Szczygiel, D., Pilecki, B., and Pojmanski, G., 2006, MNRAS, 
368, 1311

\bibitem[Perryman et al.(1997)]{Perryman97}
Perryman, M.A.C., et al., 1997, A\&A, 323L, 49 

\bibitem[Rucinski(1982)]{Rucinski82}
Rucinski, S.M., 1982, A\&A, 112, 273

\bibitem[Rucinski(1986)]{Rucinski86}
Rucinski, S.M., 1986, IAU Symp., 118, 159

\bibitem[Rucinski(2002)]{Rucinski02}
Rucinski, S.M., 2002, PASP, 114, 1124

\bibitem[Rucinski(2006)]{Rucinski06}
Rucinski, S.M., 2006, MNRAS, 368, 1319

\bibitem[Roxburgh(1966)]{Roxburgh66}
Roxburgh, I.W., 1966, ApJ, 143, 111

\bibitem[Schatzman(1959)]{Schatzman59}
Schatzman, E., 1959, A\&AS, 8, 129

\bibitem[Skumanich(1972)]{Skumanich72}
Skumanich, A., 1972, ApJ, 171, 565 

\bibitem[Stepien(1995)]{Stepien95}
Stepien, K., 1995, MNRAS, 274, 1019

\bibitem[van't Veer(1979)]{van'tVeer79}
van't Veer, F., 1979, A\&A, 80, 287

\bibitem[Vilhu \& Rahunen(1980)]{Vilhu80}
Vilhu, O., Rahunen, T., 1980, in Plavec M.J., Popper D.M., Ulrich D.R., 
eds, Proc. IAU Symp. 88, Close Binary stars, Reidel, Dordrecht, p. 491

\bibitem[Vilhu(1981)]{Vilhu81}
Vilhu, O., 1981, ApSS, 78, 401
}
\end{thebibliography}
\end{document}